\documentclass[a4paper,twosides]{article}
\usepackage{CJK,multicol,multirow,graphicx,fancyhdr,epstopdf}
\usepackage{amsmath,amsfonts,amssymb,bm,upgreek,mathrsfs,ccmap,mathcomp}
\usepackage[pagewise,switch,columnwise]{lineno}
\usepackage[compress,nospace]{cite}
\usepackage[dvipsnames]{xcolor}
\usepackage{CPL-2023}

\renewcommand{\arraystretch}{1.1} 

\newcommand{\cplyear}{2026} \newcommand{\cplvol}{43}
\newcommand{\cplno}{1} \newcommand{\cplpagenumber}{011101}
 \newcommand{\cplpage}{\cplpagenumber-\thepage}

\begin{document}

\begin{CJK*}{UTF8}{gbsn}\vspace* {-4mm} \begin{center}
\large\bf{\boldmath{Cosmic acceleration and the Hubble tension 

from baryon acoustic oscillation data}}
\footnotetext{\hspace*{-5.4mm}$^{*}$Corresponding authors. Email: gongyungui@nbu.edu.cn; luxuchen@nbu.edu.cn

\noindent\copyright\,{\cplyear}
\href{http://www.cps-net.org.cn}{Chinese Physical Society} and
\href{http://www.iop.org}{IOP Publishing Ltd}}
\\[5mm]
\normalsize \rm{}Xuchen Lu(路旭晨)$^{1*}$, Shengqing Gao(高盛晴)$^{2}$, Yungui Gong(龚云贵)$^{1*}$
\\[3mm]\small\sl $^{1}$Institute of Fundamental Physics and Quantum Technology, Department of Physics,\\ School of Physical Science and Technology, Ningbo University, Ningbo, Zhejiang 315211, China

$^{2}$School of Physics, Huazhong University of Science and Technology, Wuhan, Hubei 430074, China
\\[4mm]\normalsize\rm{}(Received 8 October 2025; accepted manuscript online 4 December 2025)
\end{center}
\end{CJK*}

\vskip 1.5mm

\small{\narrower We investigate the null tests of cosmic accelerated expansion by using the Baryon Acoustic Oscillation (BAO) data measured by the Dark Energy Spectroscopic Instrument (DESI)
and reconstruct the dimensionless Hubble parameter $E(z)$ from the DESI BAO Alcock-Paczynski (AP) data using Gaussian process to perform the null test.
We find strong evidence of accelerated expansion from the DESI BAO AP data.
By reconstructing the deceleration parameter $q(z)$ from the DESI BAO AP data,
we find that accelerated expansion persisted until $z \lesssim 0.7$ with a 99.7\% confidence level.
Additionally, to provide insights into the Hubble tension problem,
we propose combining the reconstructed $E(z)$ with $D_H/r_d$ data to derive the model-independent result $r_d h=99.8\pm 3.1$ Mpc.
This result is consistent with measurements from cosmic microwave background (CMB) anisotropies using the $\Lambda$CDM model.
We also propose a model-independent method for reconstructing the comoving angular diameter distance $D_M(z)$ from the distance modulus $\mu$ using SNe Ia data and combining this result with DESI BAO data of $D_M/r_d$ to constrain the value of $r_d$.
We find that the value of $r_d$ derived from this model-independent method is smaller than that obtained from CMB measurements, with a significant discrepancy of at least 4.17$\sigma$.
All the conclusions drawn in this paper are independent of cosmological models and gravitational theories.

\par}\vskip 3mm


\normalsize\noindent{\narrower{DOI: \href{http://dx.doi.org/10.1088/0256-307X/\cplvol/\cplno/\cplpagenumber}{10.1088/0256-307X/\cplvol/\cplno/\cplpagenumber}}
\quad \quad \quad \quad \quad
\narrower{CSTR: \href{https://cpl.iphy.ac.cn/article/cstr/32039.14.0256-307X.43.1.011101}{32039.14.0256-307X.43.1.011101}}
\par}\vskip 5mm

\begin{multicols}{2}


In 1998, observations of type Ia supernovae (SNe Ia) discovered that the Universe is experiencing accelerated expansion.\ucite{SupernovaSearchTeam:1998fmf,SupernovaCosmologyProject:1998vns}
To introduce an effective gravitational repulsive force that explains the current accelerated expansion,
the cosmological constant $\Lambda$ is one of the simplest choices.
The model that includes $5\%$ baryonic matter, $25\%$ cold dark matter (CDM), $70\%$
cosmological constant, as well as small contributions from massive neutrinos and radiation, is called the $\Lambda$CDM model.
Although the $\Lambda$CDM model is consistent with observations,
its theoretical estimation of vacuum energy is larger than the astronomical measurements by many orders of magnitude,\ucite{Weinberg:1988cp}
and it faces the fine tuning and coincidence problems.
In addition to these problems, there are other issues such as the Hubble tension.
There is a $\sim 6\sigma$ discrepancy between the Hubble constant derived from early Universe probes of cosmic microwave background (CMB)\ucite{Planck:2018vyg} and that from local measurements obtained by SNe Ia observations.\ucite{Riess:2022mme,Riess:2025chq}
The CMB measurement of the Hubble constant depends on the $\Lambda$CDM model.
The local measurements from low-redshift SNe Ia are independent of cosmological models,
but they suffer the zero-point calibration problem.

In addition to the SNe Ia and CMB data, the measurement of baryon acoustic oscillations (BAO) also supports a positive cosmological constant.
Recently, Data Release 1 (DR1) from the first year of observations by the Dark Energy Spectroscopic Instrument (DESI) gives $\Omega_\Lambda=0.651^{+0.068}_{-0.057}$ for the $\Lambda$CDM model with free spatial curvature.\ucite{DESI:2024mwx}
The result  provides strong evidence for a positive cosmological constant and accelerated expansion.
Furthermore, for the flat Chevallier-Polarski-Linder (CPL) model,\ucite{Chevallier:2000qy,Linder:2002et} DESI DR1 BAO data gives $w_0=-0.55^{+0.39}_{-0.21}$ at the $1\sigma$ confidence level and $w_a<-1.32$ at the 95\% confidence level,\ucite{DESI:2024mwx} which is in  mild tension with the $\Lambda$CDM model and suggests a dynamical dark energy;
the combination of DESI BAO data and the CMB data measured by {\it Planck}\ucite{Planck:2018vyg}
gives $w_0=-0.45^{+0.34}_{-0.21}$
and $w_a=-1.79^{+0.48}_{-1.0}$,
and the result prefers dynamical dark energy at the $\sim 2.6\sigma$ significance level;\ucite{DESI:2024mwx}
the CPL model is favored over the $\Lambda$CDM model at the $2.5\sigma$, $3.5\sigma$ and $3.9\sigma$ significance level by the combination of DESI BAO, CMB and Pantheon Plus SNe Ia,\ucite{Scolnic:2021amr} the combination of DESI BAO, CMB and Union3 SNe Ia,\ucite{Rubin:2023ovl} and the combination of DESI BAO, CMB and SNe Ia data discovered and measured during the full five year of the Dark Energy Survey (DES) program (DESY5 SNe Ia),\ucite{DES:2024jxu} respectively.\ucite{DESI:2024mwx}
With other two-parameter parametrizations of the equation of state of dark energy, a preference for a dynamical evolution towards the phantom regime  was found in.\ucite{Giare:2024gpk}
The evidence for dynamical dark energy is also supported by BAO data from other surveys.\ucite{Chan-GyungPark:2024spk}
However, the evidence for accelerated expansion or dynamical evolution of the equation of state of dark energy depends on cosmological models such as the $\Lambda$CDM model or the CPL model.

Due to model dependence,
the evidence for accelerated expansion and dynamical dark energy provided by the DESI BAO data may be biased,
and the conclusions may heavily rely on the assumptions and limitations of the chosen model.
Given the challenges associated with cosmological models,
it is crucial to explore model-independent methods for studying cosmic accelerated expansion.
Numerous parametric and non-parametric model-independent methods have been proposed to investigate energy components and the geometry of the Universe.\ucite{Visser:1997qk,Visser:1997tq,Santos:2006ja,Santos:2007pp,Gong:2007fm,Gong:2007zf,Gong:2006tx,Gong:2006gs,Seikel:2007pk,Clarkson:2007pz,Sahni:2008xx,Zunckel:2008ti,Nesseris:2010ep,Clarkson:2010bm,Shafieloo:2012ht,Holsclaw:2010nb,Holsclaw:2010sk,Holsclaw:2011wi,Seikel:2012uu,Shafieloo:2012rs,Gao:2012ef,Gong:2013bn,Yahya:2013xma,Sahni:2014ooa,Nesseris:2014mfa,Cai:2015pia,Li:2015nta,Vitenti:2015aaa,Zhang:2016tto,Wei:2016xti,Yu:2017iju,Yennapureddy:2017vvb,Velten:2017ire,Marra:2017pst,Melia:2018tzi,Gomez-Valent:2018hwc,Pinho:2018unz,Haridasu:2018gqm,Capozziello:2018jya,Arjona:2019fwb,Jesus:2019nnk,Franco:2019wbj,Bengaly:2019ibu,Yang:2019fjt,Dhawan:2021mel,Gangopadhyay:2023nli,Sharma:2024mtq,Dinda:2024kjf,Jiang:2024xnu,Ghosh:2024kyd,Cortes:2024lgw,Shlivko:2024llw,deCruzPerez:2024shj,Roy:2024kni,Chatrchyan:2024xjj,Perivolaropoulos:2024yxv,Linder:2024rdj,Payeur:2024dnq,Chan-GyungPark:2024brx,Gao:2024ily,Gao:2025ozb,Dinda:2025hiu}
The purpose of this work is to present agnostic evidence for accelerated expansion.

We start with the following two conditions
\begin{equation}\label{eq:q}
    q(t)=-\frac{\ddot{a}} {\left(a H^2\right)} \geq 0,
\end{equation}
\begin{equation}\label{eq:doth}
    \quad \dot{H}-\frac{k}{a^2} \leq 0,
\end{equation}
where Eq. \eqref{eq:q} indicates decelerated expansion,
and Eq. \eqref{eq:doth} indicates no super-accelerated expansion.
We refer to these two conditions as the null test conditions.
It is important to note that the null test conditions depend solely on the assumption of the Friedmann-Robertson-Walker metric which is based on the cosmological principle, and are independent of any specific gravity theory or cosmological model.

For Eq. \eqref{eq:q},
the deceleration parameter $q$ can be reformulated as a function of redshift $z$
\begin{equation}\label{eq:qz}
    q=-\frac{\ddot{a}}{a H^2}=\frac{H^{\prime}}{H}(1+z)-1,
\end{equation}
where $^\prime$ denotes the derivative with respect to $z$.
Subsequently, Eq. \eqref{eq:qz} can be expressed in the following integral form
\begin{equation}\label{eq:intergral}
    \ln \frac{H(z)}{H_0}=\int_0^z \frac{1+q\left(z^{\prime}\right)}{1+z^{\prime}} d z^{\prime} .
\end{equation}
Substituting Eq. \eqref{eq:q} into Eq. \eqref{eq:intergral},
we obtain
\begin{equation}
\label{eq:Hzgeq}
    H(z)\geq H_0(1+z).
\end{equation}
If the Universe has never experienced accelerated expansion,
or has always remained in a state of decelerated expansion, Eq. \eqref{eq:Hzgeq} will hold.
This argument provides a model-independent method to determine whether the Universe is undergoing accelerated expansion.
However, based on the discussion in Refs. \cite{Gong:2007fm,Gong:2007zf},
it is important to emphasize that the transition from Eq. \eqref{eq:q} to Eq. \eqref{eq:Hzgeq} involves an integration.
In other words, even if Eq. \eqref{eq:Hzgeq} holds over specific redshift ranges,
it does not imply that the Universe has never experienced accelerated expansion within those intervals.
Similarly, even if Eq. \eqref{eq:Hzgeq} is violated within a certain redshift interval,
it does not mean that the Universe was always in a state of accelerated expansion throughout that interval.
Conversely, if Eq. \eqref{eq:Hzgeq} is violated at certain redshifts, it provides conclusive evidence that the Universe has undergone a phase of accelerated expansion.

Similarly, from Eq. \eqref{eq:doth},
we obtain
\begin{equation}\label{eq:HzOk}
    H(z)\geq H_0\sqrt{1-\Omega_{k0}+\Omega_{k0}(1+z)^2},
\end{equation}
where $\Omega_{k0}=-k/(a_0H_0^2)$ is the curvature density parameter.
As with the discussion of Eq. \eqref{eq:Hzgeq},
if Eq. \eqref{eq:HzOk} is not satisfied at a certain redshift,
we can infer that the Universe has undergone a phase of super-accelerated expansion, which means a deviation from the $\Lambda$CDM model in the context of general relativity.

Note that for redshifts $z\geq 0$ , we naturally obtain
\begin{equation}\label{eq:HzOk1}
    H_0(1+z)\geq H_0\sqrt{1-\Omega_{k0}+\Omega_{k0}(1+z)^2}.
\end{equation}
Thus, satisfying Eq. \eqref{eq:HzOk} guarantees the satisfaction of Eq. \eqref{eq:Hzgeq}.
In other words, if the Universe has undergone super-accelerated expansion,
it must have also experienced accelerated expansion.

For a flat universe with $\Omega_{k0}=0$, Eq. \eqref{eq:HzOk} simplifies to
\begin{equation}
\label{eq:Hzgeq2}
 H(z)\geq H_0.
\end{equation}

The DESI BAO data gives measurements on the quantities $D_M(z)/r_d$ and $D_H(z)/r_d$,
where $D_M(z)$ is the transverse comoving distance, $r_d$ is the drag-epoch sound horizon,
and the line-of-sight distance variable
\begin{equation}
\label{eq:D_H}
    D_H=\frac{c}{H(z)}.
\end{equation}
where $c$ is the speed of light.
Assuming the Universe is homogeneous and isotropic on large scales,
the spacetime geometry is described by the Friedmann-Robertson-Walker metric, and the transverse comoving distance $D_{M}$ is given by
\begin{equation}
\label{eq:D_M}
    D_{M}(z)=\frac{c}{H_0 \sqrt{|\Omega_{k0}|}} \text{sinn} \left[\sqrt{|\Omega_{k0}|} \int_0^z \frac{d z^{\prime}}{E\left(z^{\prime}\right)}\right],
\end{equation}
where $E(z)=H(z)/H_0$ is the dimensionless Hubble parameter, $H_0$ is the Hubble constant
and $\text{sinn}(\sqrt{|\Omega_{k0}|}x)/\sqrt{|\Omega_{k0}|}=\sinh(\sqrt{|\Omega_{k0}|}x)/\sqrt{|\Omega_{k0}|}$, $x$,
and $\sin(\sqrt{|\Omega_{k0}|}x)/\sqrt{|\Omega_{k0}|}$ for $\Omega_{k0}>0$,
$\Omega_{k0}=0$ and $\Omega_{k0}<0$, respectively.
For the flat universe, $\Omega_{k0}=0$,
\begin{equation}
\label{eq:D_M_flat}
D_M(z)=\frac{c}{H_0}\int_{0}^{z}\frac{d z^{\prime}}{E\left(z^{\prime}\right)}.
\end{equation}
The sound horizon at the drag epoch $z_d$ is
\begin{equation}
\label{eq:r_d}
r_d=\int_{z_d}^\infty \frac{c_s(z)}{H(z)}dz,
\end{equation}
where $c_s(z)$ is the speed of sound in the baryon-photon plasma.
The DESI BAO measurements used in our analysis are listed in Table \fref{table:1}{1}.

\vskip 4mm

\tl{table:1}\tabtitle{7.5}{1}{The DESI DR1 BAO measurements.
For each tracer, we quote the effective redshift $z_{\rm eff}$, $D_M/r_d$, $D_H/r_d$ and the correlation coefficient $r$. 
This table can be found in Ref. \cite{DESI:2024mwx}.}

\vskip 2mm
\tabcolsep 3pt
\renewcommand{\arraystretch}{1.5}

\centerline{\footnotesize
\begin{tabular}{ccccc}
\hline\hline\hline
Tracer & $z_{\rm eff}$ & $D_M/r_d$ & $D_H/r_d$ & $r$ \\
\hline
LRG1 & 0.510 & $13.62 \pm 0.25$ & $20.98 \pm 0.61$ & $-0.445$ \\
LRG2 & 0.706 & $16.85 \pm 0.32$ & $20.08 \pm 0.60$ & $-0.420$ \\
LRG3+ELG1 & 0.930 & $21.71 \pm 0.28$ & $17.88 \pm 0.35$ & $-0.389$ \\
ELG2 & 1.317 & $27.79 \pm 0.69$ & $13.82 \pm 0.42$ & $-0.444$ \\
Lya QSO & 2.330 & $39.71 \pm 0.94$ & $8.52 \pm 0.17$ & $-0.477$ \\
\hline\hline\hline
\end{tabular}}

\vskip 4mm

Note that the Alcock-Paczynski (AP) parameter, defined as $F_{AP}=D_M/D_H$,
is independent of the cosmological parameter $r_d$.
This property of $F_{AP}$ was used to perform the null test of cosmic curvature and discuss the compatibility between DESI BAO and SNe Ia data.\ucite{Gao:2025ozb,Dinda:2025hiu}
To use $F_{AP}$ from the measurements of $D_M/r_d$ or $D_H/r_d$,
it is necessary to properly account for the correlation between these two quantities.
Therefore, we derive $F_{AP}$ and its associated uncertainty by using both
$D_M/r_d$, $D_H/r_d$, and their covariance matrix as provided in Table \fref{table:1}{1}.

In a flat universe, $F_{AP}$ is
\begin{equation}\label{eq:fap_ez}
F_{AP}=E(z)\int_0^z\frac{1}{E(z^\prime)}dz^\prime,
\end{equation}
so for a spatially flat universe,
\begin{equation}\label{eq:FapEz}
E(z)=\exp{\int_0^z\frac{F_{AP}^\prime(z^\prime)-1}{F_{AP}(z^\prime)}dz^\prime}.
\end{equation}
In the following discussion,
we assume a spatially flat universe.
To analyze the data, 
we employ the Gaussian Process (GP), a robust statistical technique that models distributions over functions within a Bayesian framework.\ucite{Holsclaw:2010nb,Holsclaw:2010sk,Holsclaw:2011wi,Seikel:2012uu,Cai:2016vmn,Miao:2015dua,Yang:2015tzc}
By utilizing Gaussian distributions and a covariance (or kernel) function, 
this method effectively captures the correlations between data points.
We choose the squared exponential kernel function. 
In general, the reconstructed results depend on kernel functions,\ucite{Johnson:2025blf} 
but our reconstructed results from $F_{AP}$ are robust to the specific kernel choice.
Applying this GP method to the DESI BAO AP data, we reconstruct $E(z)$ and $F_{AP}$.
The reconstructed $E(z)$ are shown in Figs. \fref{fig:Ez}{1} and \fref{fig:Ez1}{2}.

Since the first point at the redshift $z=0.51$ exhibits a statistical fluctuation,\ucite{Colgain:2024xqj,DESI:2024mwx}
we remove the point to perform the GP reconstruction again, and labeled the results as DESI BAO$^{-}$.
From Fig. \fref{fig:Ez}{1}, we see that the null condition \eqref{eq:Hzgeq} is violated up to $z\lesssim 1.2$ with more than $1\sigma$ significance,
and the null condition \eqref{eq:Hzgeq2} is always satisfied.
Without the data point $F_{AP}(z=0.51)$, even though the error bar of $E(z)$ is larger, but the evidence of accelerated expansion is stronger.
Therefore, the DESI BAO AP data provides both strong evidences of accelerated expansion and past decelerated expansion, but there is no evidence of super-accelerated expansion if we assume a spatially flat universe.
\vskip 4mm

\fl{fig:Ez}\centerline{\includegraphics[width=0.6\linewidth]{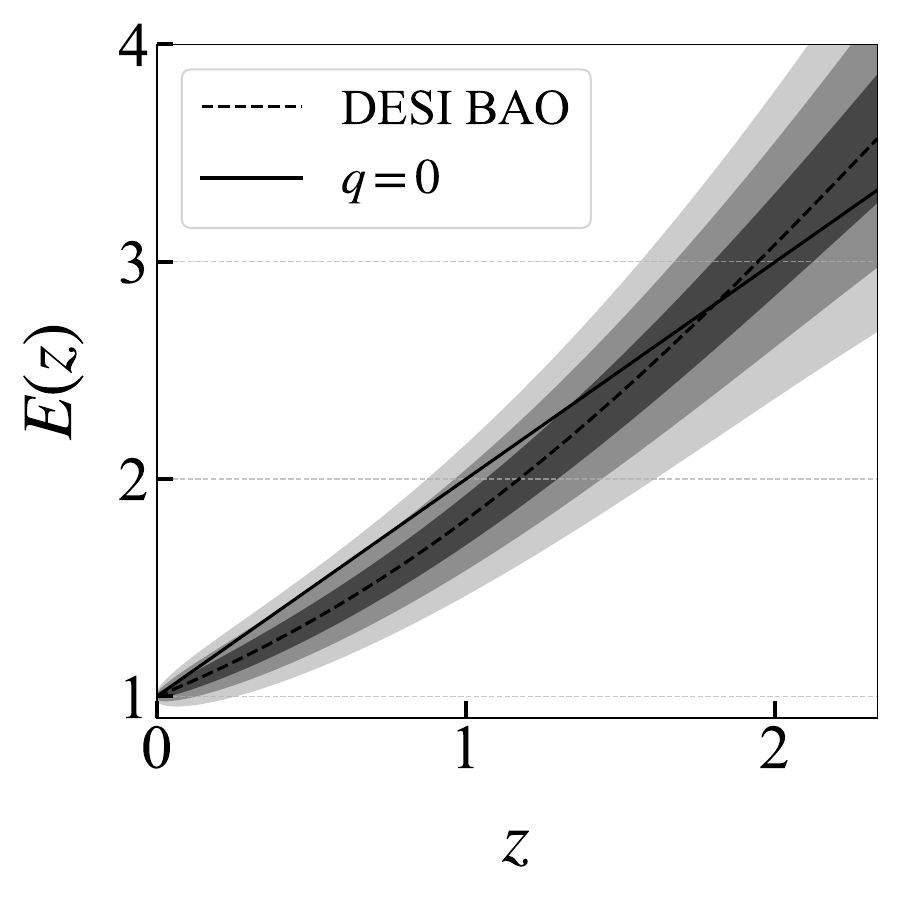}}
\vskip 2mm

\figcaption{7.5}{1}{The GP reconstructed $E(z)$ using the DESI BAO AP data.
The dashed grey line represents the reconstructed result,
with shaded grey region indicating the 1$\sigma$, 2$\sigma$ and 3$\sigma$ uncertainty.
The solid black line corresponds to the null condition $q(z)=0$.}

\medskip

\vskip 4mm

\fl{fig:Ez1}\centerline{\includegraphics[width=0.6\linewidth]{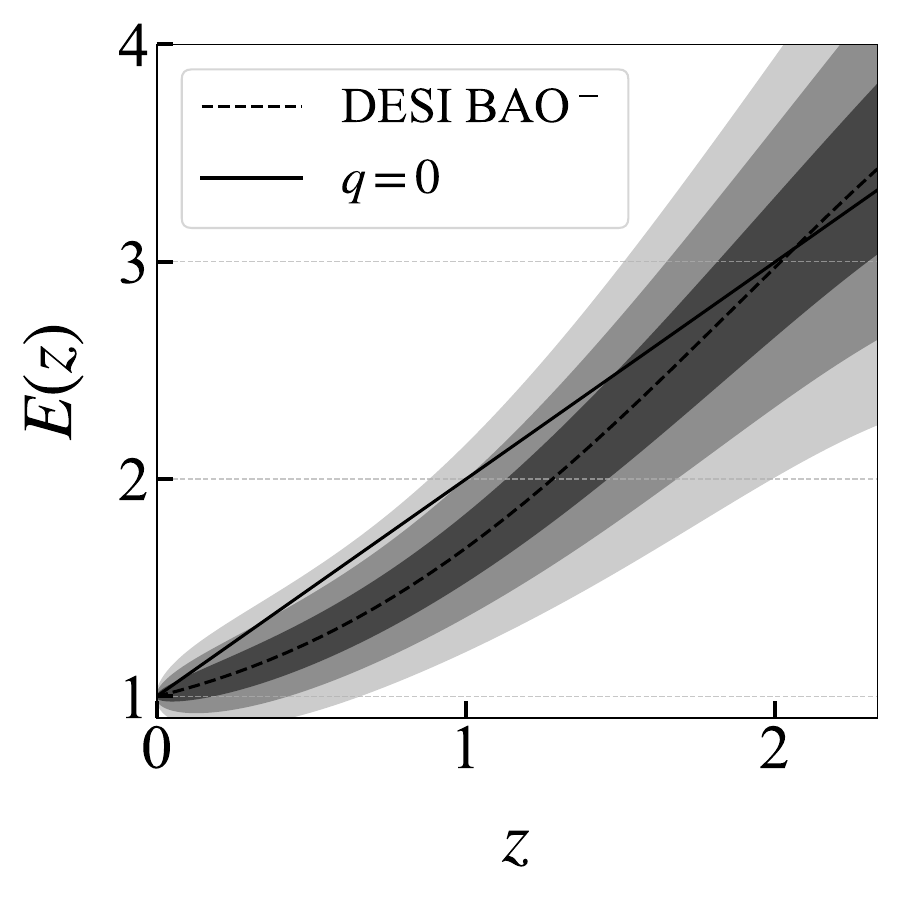}}
\vskip 2mm

\figcaption{7.5}{2}{The GP reconstructed $E(z)$ using the DESI BAO$^{-}$ data. 
The lines and shaded regions are the same as in Fig. \fref{fig:Ez}{1}}

\medskip

\vskip 4mm

\fl{fig:nz}\centerline{\includegraphics[width=0.6\linewidth]{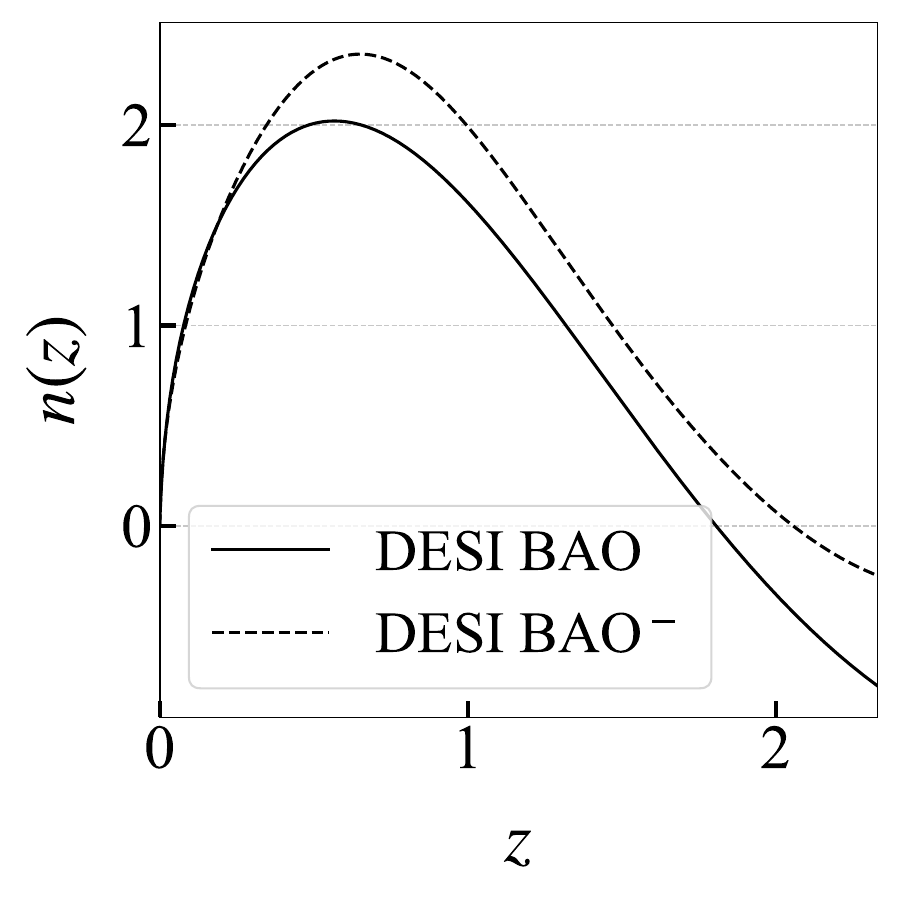}}
\vskip 2mm

\figcaption{7.5}{3}{The results of $n(z)$ using the DESI BAO AP data. We label the datasets without the data point $F_{AP}(z=0.51)$ as DESI BAO$^{-}$.}

\medskip

To quantify the evidence provided by the null test with reconstructed $E(z)$,
we use the function $n(z)$\ucite{Seikel:2007pk},
\begin{equation}
    n(z) = \frac{(1+z)-E(z)}{\sigma(z)},
\end{equation}
to assess the significance interval of the deviation from the null hypothesis,
where $\sigma(z)$ is the error of the GP reconstructed $E(z)$.
For example, $n(z)=2.326$ means a 99\% confidence level and $n(z)=1.645$ means a 95\% confidence level at the redshift $z$.
In Fig. \fref{fig:nz}{3}, we plot the results of $n(z)$ in the redshift interval $[0, 2.33]$, for cases with and without the data point $F_{AP}(z=0.51)$.
From Fig. \fref{fig:nz}{3}, we see that $n(z)$ reaches the maximum at around $z\sim 0.7$ and the maximum value of $n(z)$ is larger than 2,
indicating very strong evidence for cosmic acceleration and the transition redshift $z\sim 0.7$ at which the transition from accelerated to decelerated expansion happened.

To further explore the model-independent evidence for accelerated expansion using the DESI BAO data,
we discuss the reconstruction of the deceleration parameter $q(z)$ from the $F_{AP}$ data.
The deceleration parameter $q(z)$ can be related to $E(z)$ as follows,
\begin{equation}\label{eq:qz_ez}
q(z)=\frac{E^\prime(z)}{E(z)}(1+z)-1.
\end{equation}
By combining Eqs. $\eqref{eq:fap_ez}$ and $\eqref{eq:qz_ez}$, we arrive at
\begin{equation}\label{eq:qz_fap}
q(z)=\frac{F^\prime_{AP}-1}{F_{AP}}(1+z)-1,
\end{equation}
which provides a direct relationship between the AP parameter $F_{AP}$ and the deceleration parameter,
offering an insightful way to probe the expansion history of the Universe using the DESI BAO data.
The reconstructed $q(z)$ as shown in Figs. \fref{fig:qz}{4} and \fref{fig:qz}{5}.

From Figs. \fref{fig:qz}{4} and \fref{fig:qz}{5}, we observe that the accelerated expansion persisted until $z \lesssim 0.7$ with a 99.7\% confidence level,
whereas for the $\Lambda$CDM model, the transition from accelerated expansion to decelerated expansion happened at $z \simeq 0.689$.
When the data point $F_{AP}(z=0.51)$ is excluded,
even though the uncertainty in the GP reconstructed $q(z)$ increases significantly at low redshifts,
the more than $3\sigma$ evidence for accelerated expansion persists.

The local measurement of the Hubble constant using SNe Ia calibrated with Cepheids by SH0ES is
$H_0=73.18\pm 0.88$ km s$^{-1}$Mpc$^{-1}$.\ucite{Riess:2025chq}
This value is in $\sim 6\sigma$ tension with the measured value $H_0=67.36\pm 0.54$ km s$^{-1}$Mpc$^{-1}$ from the Planck 2018 data using the $\Lambda$CDM model.\ucite{Planck:2018vyg}
Based on the flat $\Lambda$CDM model, the CMB and CMB lensing data yield $r_d = 147.09 \pm 0.26$ Mpc  and $r_d h=98.82\pm 0.82$ Mpc,\ucite{Planck:2018vyg}
while the DESI BAO data provide $r_d h=101.8\pm 1.3$ Mpc,\ucite{DESI:2024mwx}
so the two measurements from early and late universe on $r_dh$ exhibit mild tension.
The tension increases with the model-independent value $r_dh=104.02 \pm 2.34$ Mpc obtained from the multi-task GP reconstruction of $D_H(z)/r_d$.\ucite{Dinda:2024ktd}
If the value of the Hubble constant measured by the CMB increases while the value of $r_d$ measured by the CMB remains the same, then the value of $r_d h$ increases and the tension is alleviated.
Combining the CMB measurement on $r_d$ and BAO measurement on $r_d h$, one gets $H_0=69.29\pm 0.87$ km s$^{-1}$Mpc$^{-1}$,\ucite{DESI:2024mwx}
which alleviates the tension somewhat.
We may argue that the low redshift BAO data contributes to reducing the tension.
However, the result depends on the $\Lambda$CDM model and a significant factor contributing to the Hubble tension could be the discrepancy between the value of $r_d$ derived from early universe observations and local measurements.\ucite{Perivolaropoulos:2024yxv}

\vskip 4mm

\fl{fig:qz}\centerline{\includegraphics[width=0.6\linewidth]{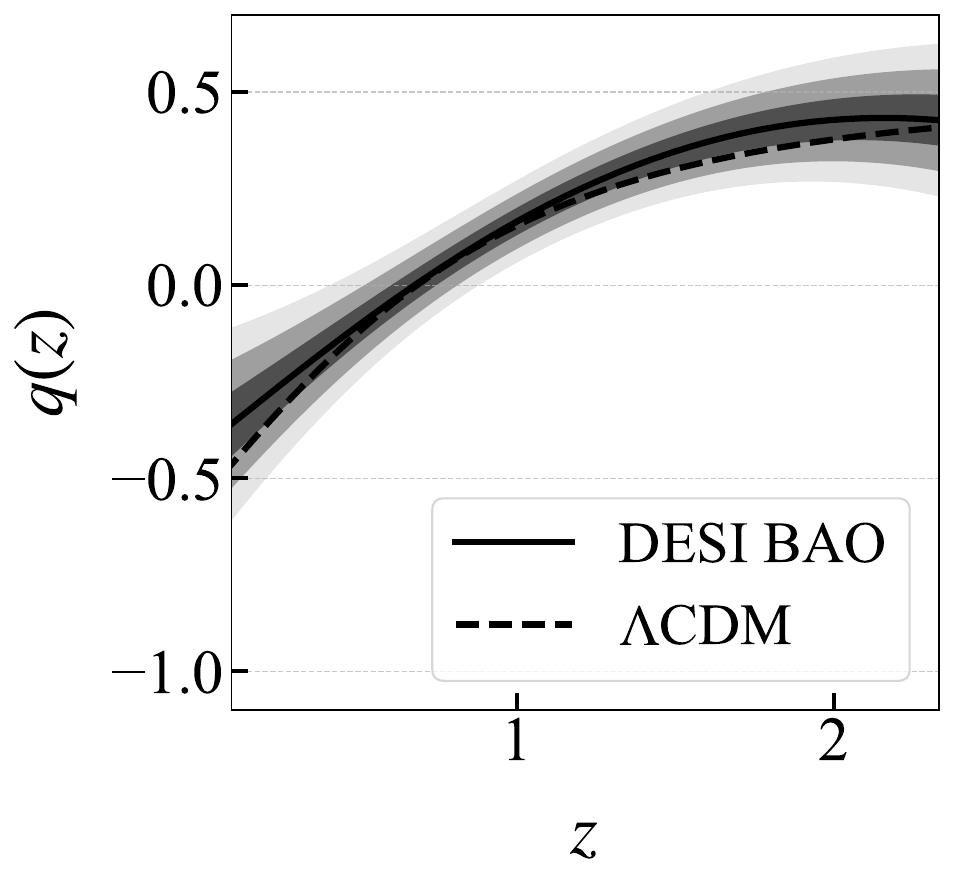}}
\vskip 2mm

\figcaption{7.5}{4}{
The reconstructed deceleration parameter $q(z)$ using the DESI BAO $F_{AP}$ data.
The solid lines represent the reconstructed results,
with shaded regions indicating the associated $1\sigma$, $2\sigma$ and $3\sigma$ uncertainties.
For comparison, the dashed lines show the result for the $\Lambda$CDM model with $\Omega_{m0}=0.295$.}
\medskip

\vskip 4mm

\fl{fig:qz1}\centerline{\includegraphics[width=0.6\linewidth]{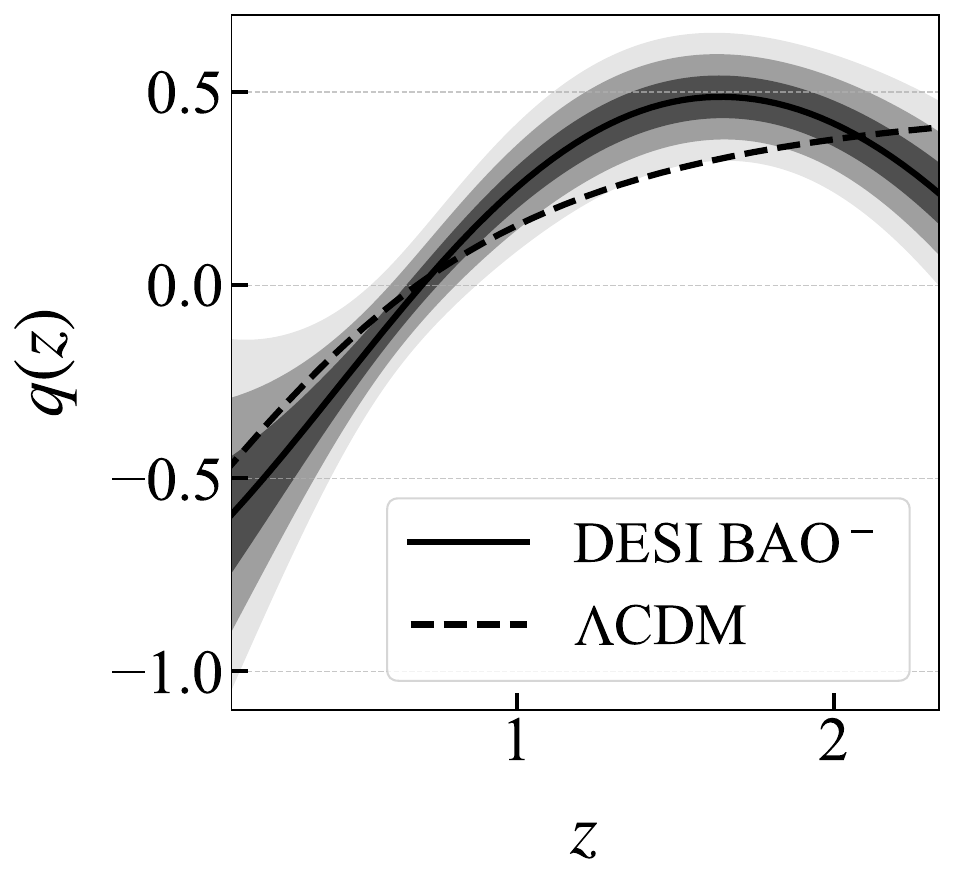}}
\vskip 2mm

\figcaption{7.5}{5}{The reconstructed deceleration parameter $q(z)$ using the DESI BAO$^{-}$ data. 
The lines and shaded regions are the same as in Fig. \fref{fig:4}{4}.}
\medskip

To explore the Hubble tension problem with the DESI BAO data, we combine the reconstructed $E(z)$ as shown in Figs. \fref{fig:Ez}{1} and \fref{fig:Ez1}{2} with the $D_{H}(z)/r_d$ data to provide a model-independent constraint on $r_dh$ because $\tilde{D}_H=D_{H}(z)/r_d=c/[H_0 E(z)r_d]$.
To constrain the value of $r_d h$, we minimize
\begin{equation}
\chi^2=\sum_{i}^{N}\left[\frac{\left[D_H(z_i)/r_d-\tilde{D}_H(z_i)\right]^2}{(\sigma_{D_H/r_d}^i)^2+(\sigma^i_R)^2}\right]
\end{equation}
with iminuit,\ucite{iminuit}
where $\sigma_{D_H/r_d}^i$ is the error of the data, $\sigma^i_R$ is the error from GP reconstruction of $E(z)$, $N$ is the number of DESI BAO data $D_H/r_d$ we used.
Minimizing this $\chi^2$ function yields the 1$\sigma$ constraint
$r_d h=99.7\pm 3.1$ Mpc,
and the corresponding posterior probability distribution is shown in Fig. \fref{fig:rdh}{6}.
The model-independent result is consistent with both CMB and DESI BAO results using the $\Lambda$CDM model,
indicating that there is no tension between the DESI BAO and Planck data.

\vskip 4mm

\fl{fig:rdh}\centerline{\includegraphics[width=0.7\linewidth]{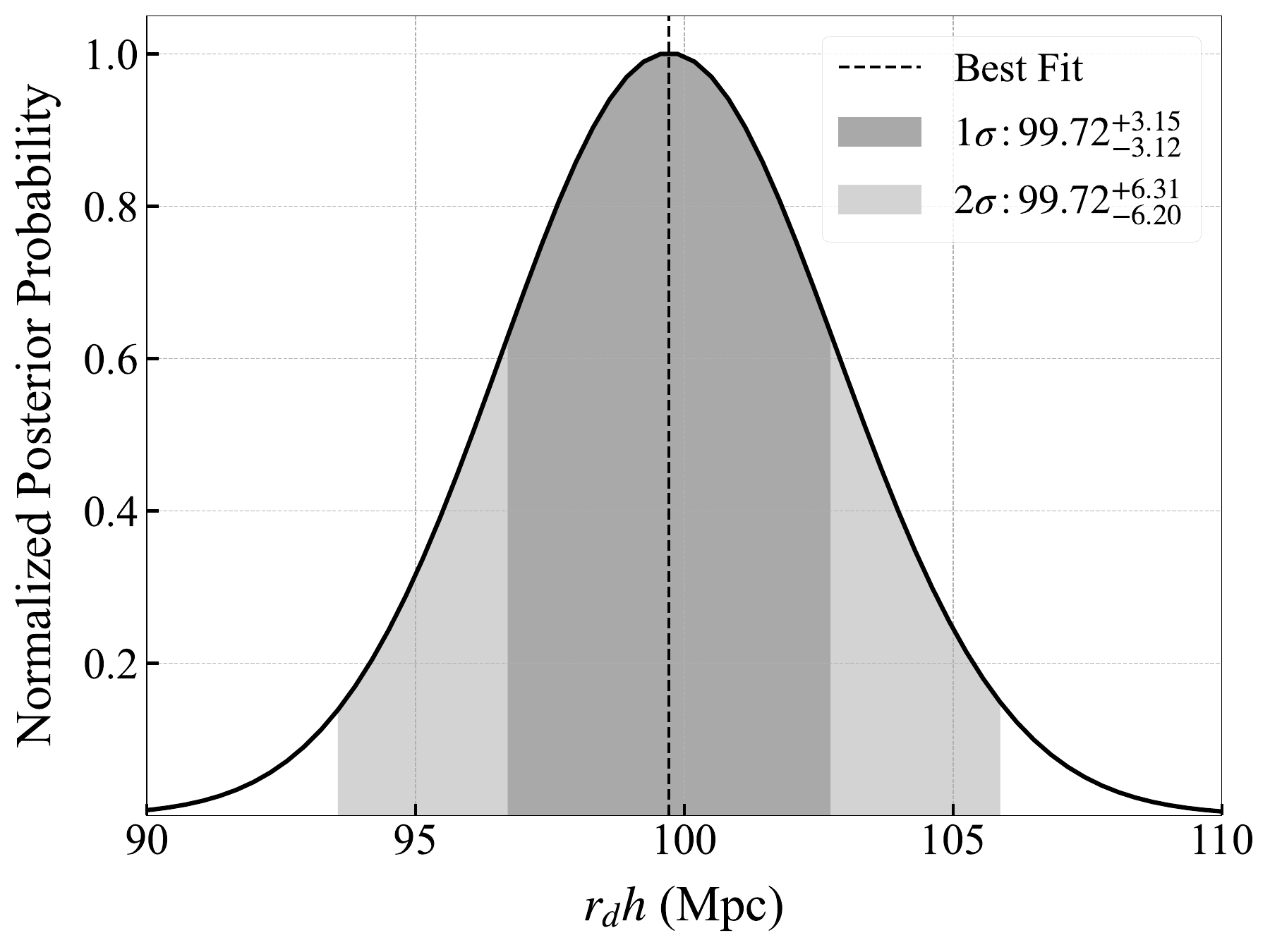}}
\vskip 2mm

\figcaption{7.5}{6}{The normalized posterior probability distribution for $r_dh$.The vertical dashed line marks the best-fit value, while the dark and light gray shaded regions show the 1$\sigma$ and 2$\sigma$ confidence regions, respectively.}

\medskip

Since DESI BAO data provide $D_M(z)/r_d$, and the luminosity distance can be measured by SNe Ia observations,
it is feasible to derive a model-independent value of $r_d$ by combining the DESI BAO data and SNe Ia data.
This allows us to compare this model-independent determination of $r_d$ with the CMB result to investigate the Hubble tension problem.

In a homogeneous and isotropic universe, the transverse comoving distance $D_M$ is related to the luminosity distance $d_L$ as
\begin{equation}
d_L(z)=(1+z)D_{M}(z).
\end{equation}
SNe Ia data provide the distance modulus $\mu(z)$, which is related to the luminosity distance $d_L(z)$ as
\begin{equation}
\mu=5\log_{10}\left[\frac{d_L(z)}{\text{Mpc}} \right]+25.
\end{equation}

We reconstruct the distance modulus $\mu(z)$ from SNe Ia data by using the GP method, yielding the reconstructed $d_L(z)$ and its associated error $\delta d_L(z)$.
We then define the function $\tilde{D}(z)=d_L(z)/r_d(1+z)$,
with $r_d$ as a free parameter to fit the DESI BAO data $D_M(z)/r_d$.
To constrain the value of $r_d$, we minimize
\begin{equation}
\chi^2=\sum_{i}^{N}\left[\frac{\left[(D_M/r_d)_i-\tilde{D}(z_i)\right]^2}{(\sigma_{D_M/r_d}^i)^2+(\sigma^i_P)^2}\right]
\end{equation}
with iminuit,
where $\sigma^i_P=\delta d_L(z_i)/r_d(1+z_i)$ is the error from GP reconstruction.
The SNe Ia datasets considered in this paper include Union3.0,\ucite{Rubin:2023ovl} Pantheon Plus,\ucite{Scolnic:2021amr} and DESY5.\ucite{DES:2024jxu}
The results for $r_d$ and the corresponding uncertainties derived from these different SN Ia datasets are shown in Fig. \fref{fig:rd_tension}{7}.

\vskip 4mm

\fl{fig:rd_tension}\centerline{\includegraphics[width=0.85\linewidth]{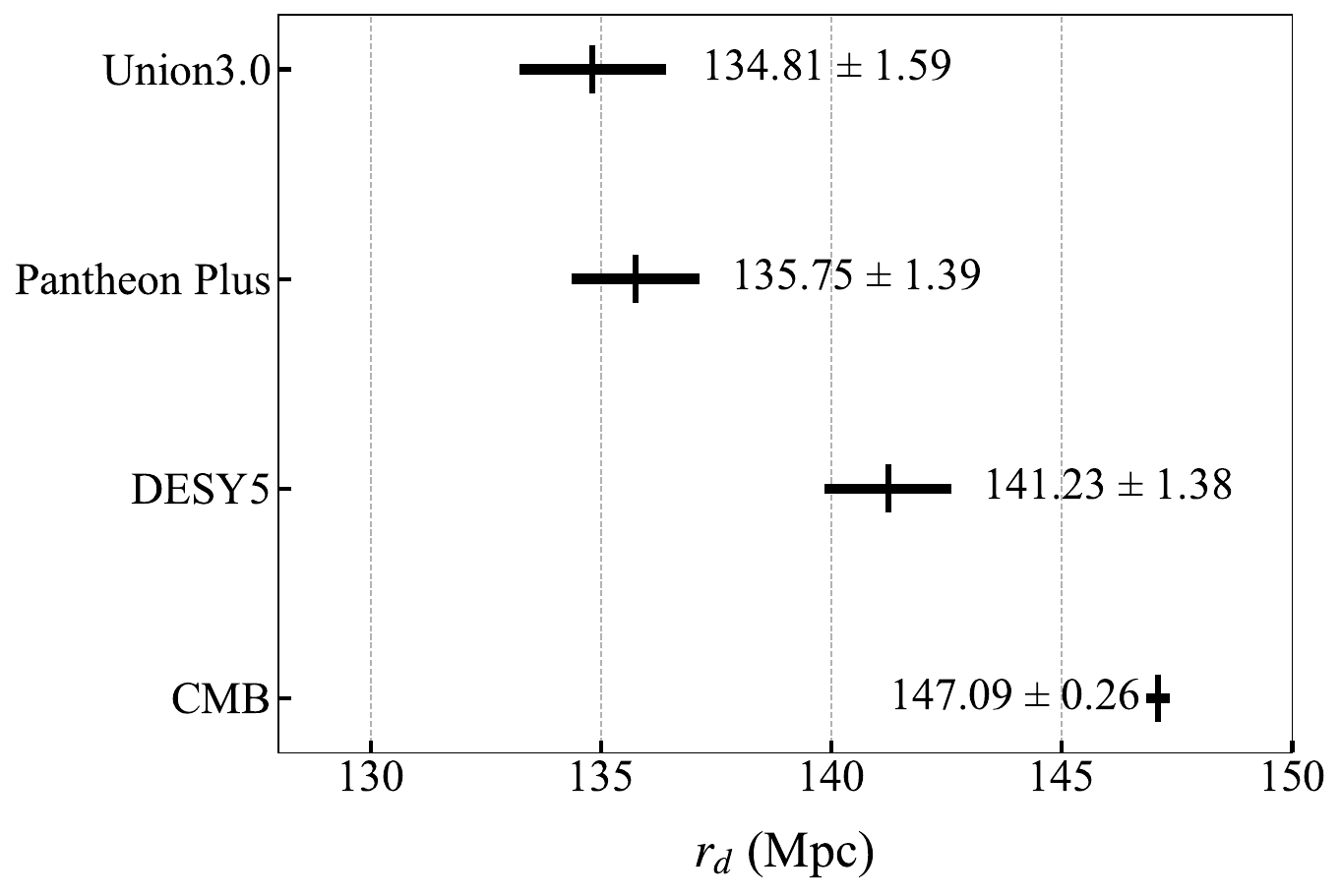}}
\vskip 2mm

\figcaption{7.5}{7}{Measurement of $r_d$ from the combination of DESI BAO and various SNe Ia datasets.}

\medskip

From Fig. \fref{fig:rd_tension}{7}, we see that the values of $r_d$ constrained by SNe Ia data are consistently smaller than that given by the CMB data.
The values of $r_d$ derived from the datasets combined with Union3, Pantheon, and DESY5  differ from the CMB result by 7.62$\sigma$, 8.02$\sigma$, and 4.17$\sigma$, respectively.
Note that the results from the SNe Ia and DESI BAO data are independent of cosmological models, they just rely on the assumption of cosmological principle.
Since both BAO and SNe Ia data are independent of early universe physics,
the discrepancy between the value of $r_d$ derived from early universe observations and low-redshift measurements may provide hints on the Hubble tension problem.
Of course, the method also suffers the same zero-point calibration issue as SNe Ia data.

In summary, we use DESI BAO data to perform a null hypothesis test on cosmic accelerated expansion,
reconstruct the dimensionless Hubble parameter $E(z)$ and the deceleration parameter $q(z)$,
determine the value of $r_d h$ from DESI BAO data alone and the value of $r_d$ in conjunction with SNe Ia data.
We find strong evidence of accelerated expansion if we assume a spatially flat universe.
The value of $n(z)$ reaches the maximum at around $z\sim 0.7$ and the maximum value of $n(z)$ is larger than 2,
indicating that the cosmic accelerated expansion persisted up to $z\sim 0.7$.
By reconstructing the deceleration parameter $q(z)$ from the DESI BAO AP data,
we observe with a 99.7\% confidence level that accelerated expansion occurred until $z \lesssim 0.7$. The result is consistent with that derived from the null test with the resconstruced $E(z)$.

Combining the reconstructed $E(z)$ with $D_H/r_d$ data, we derive the model-independent result $r_d h=99.7 \pm 3.1$ Mpc.
The result is consistent with that obtained from measurements of CMB anisotropies using the $\Lambda$CDM model by {\it Planck},
indicating that there is no tension between the DESI BAO and Planck 2018 data.
With more data and enhanced accuracy in future observations, we believe that the error bar of $r_d h$ can be reduced,
and further insight into the Hubble tension problem can be gained.

By proposing the model-independent method of reconstructing the comoving angular diameter distance $D_M(z)$ from SNe Ia distance modulus data and combining the result with DESI BAO data of $D_M/r_d$ to constrain the value of $r_d$,
we find that the value of $r_d$ derived from this model-independent method is smaller than that obtained from CMB measurements, with a significant discrepancy of at least 4.17$\sigma$.
The tension may be caused by the zero-point calibration of SNe Ia data.
The discrepancy between the value of $r_d$ derived from early universe observations and low-redshift measurements may offer clues regarding on the Hubble tension problem.
We would like to emphasize that all the conclusions drawn are independent of cosmological models and gravitational theories.

\textit{Acknowledgements.} This work is supported in part by the National Key Research and Development Program of China under Grant No. 2020YFC2201504, 
the National Natural Science Foundation of China Grant Nos. 12588101 and 12535002.
We would like to thank Eoin Colg\'{a}in and Stefano Anselmi for fruitful comments.
We are also grateful to Filipe B. Abdalla for helpful discussions.


\end{multicols}
\end{document}